# A step towards the Analysis of the Influence
# of an electrical Field on the Propagation of Light

Wolfenbüttel, October - 8 - 2007

Claus W. Turtur, University of Applied Sciences Braunschweig-Wolfenbüttel, Germany

## Abstract:

It is a matter of common knowledge that the field of gravitation influences the propagation of light, but the question to which extent electromagnetic fields do something similar is not completely answered. The birefringence caused by electric and magnetic fields is subject of several articles in Quantumelectrodynamics, but the absolute refractive index is discussed rather rarely.
Some hypothetical considerations about the feasibility of an experiment to analyze the refractive index in electromagnetic fields were given in a former article by the same author. A first experimental test is reported now.[1] Perhaps an evidence for an influence of an electric field on the refractive index was found, but the result is not reliable enough to claim already, that the influence was seen – but it should encourage to perform further measurements with different methods of higher precision.

## Structure of the paper:

1. Basic principles
2. Experimental setup
3. Measurement, analysis and discussion of the results
4. Considerations about further investigation

## Contents:

### 1. Basic principles

The influence of the fundamental interaction of gravitation on the propagation of light is one of the subjects of the theory of General Relativity (for instance [SCH 02] or [GOE 96]). Regarding the influence of the electromagnetic interaction on the propagation of light, some theoretical considerations can be found. For instance in [RIZ 07] there is an indication regarding the birefringent properties of magnetic fields in vacuum. The value of this so called Cotton-Mouton-birefringence is given in [BIA 70] as $\Delta n_{CM} = n_{\parallel} - n_{\perp} = 4 \cdot 10^{-24} \cdot B^2$ (with $B =$ magnetic field strength in Tesla). The influence of electric and magnetic fields on light is one of the subjects of [RIK 00], who specifies the Cotton-Mouton-birefringence to be $\Delta n_{CM} = n_{\parallel} - n_{\perp} = 3.5 \cdot 10^{-21}$ at a magnetic field of 30 Tesla and a Kerr-birefringence of $\Delta n_K = n_{\parallel} - n_{\perp} = -3.4 \cdot 10^{-25}$ at an electric field of $3 \, kV/cm$. It is obvious that such small effects give a substantial challenge to any measurement.

The question is now, whether it might be possible to find any other effect, which leads to a stronger influence of electric or/and magnetic fields on light, so that the measurement would be easier. In the case of gravitation it is not birefringence (which only describes a spatial anisotropy of the refractive index), which made it possible to check the theory of general relativity, but it is the deflection of light (for instance in the gravitational field of the sun), which is nowadays understood as one of the proofs of general relativity, and which can be interpreted in the sense of the absolute refractive index. (See [KLÜ 60] with regard to visible wavelength and [FOM 76], [FOM 77] with regard to electromagnetic radiowaves.) Insofar there is well-founded hope that electric or/and magnetic fields might cause a deflection of light or may have an influence on the speed of propagation, which might be sufficient for measurement. This could be described as an alteration of the refractive index caused by a field, which means that the refractive index of the vacuum with field $n_{field} = 1 + \Delta n$ would be larger than the refractive index of the vacuum without field $n_{vacuum} = 1$. And $\Delta n$ might be within a quite different order of magnitude than in the case of birefringence.

---

[1] Special thanks go to Prof. Dr.-Ing. habil. G.-P. Ostermeyer from the Technical Universität of Braunschweig who provided his white light interferometer located in a laboratory of constant temperature and to Dipl.- Ing. J.-H. Sick who helped performing the practical measurements in the laboratory.

For the purpose of the planning of such an experiment, in [TUR 07] a rough estimation of the order of magnitude of $\Delta n$ was attempted, which intends to be nothing more than only the basis for the planning of an experiment, because of its substantial numerical uncertainties and because of the lack of a fundamental model based on a physical mechanism.

The experiment drafted there has been performed in meantime, whereas the design of the interferometer had to be adjusted to the practical circumstances in the laboratory. It is the central matter of the following pages. Thereby some indications have been achieved, that the speed of propagation of light, might be influenced by an electric field. However, the effect is such tiny, that in the moment the result should be interpreted as a qualitative hint for the existence of the influence and not yet as a quantitative specification of numerical values.

## 2. Experimental setup

The basic idea of the experiment consists of the use of an interferometer, in which two parts of a light beam are brought to interference. Let us denominate one beam to be the reference-beam which is neither exposed to a electric nor to a magnetic field. The other beam shall be the measuring-beam which can be exposed to a field optionally (see fig.1). In the experiment, an electric field was generated by the use of a capacitor, a magnetic field was not applied. The measuring-beam was reflected by a glass surface being as smooth as possible. This surface was orientated nearly but not completely perpendicular to the beam in order to produce an interference pattern consisting of few stripes, each of them representing a maximum and an minimum of interference, corresponding to a path length difference of $2 \cdot \frac{\lambda}{2} = 520\,\text{nm}$ (with the wavelength used in the interferometer sketched in fig.1).

The operation of switching-on the field $\vec{E}$ should reduce the speed of propagation of measuring-beam and by that shift the pattern of interference in the same way as a prolongation of the light path would do, and because of this it should shift the gray scale value at the accordant positions of the interference pattern, corresponding to a shift of the optical retardation. As can be seen in fig.1, the capacitor consists of tapered plates which generate an inhomogeneous field, which shall cause an inhomogeneous alteration of gray scale values – in the case that the field influences the speed of propagation of the light.

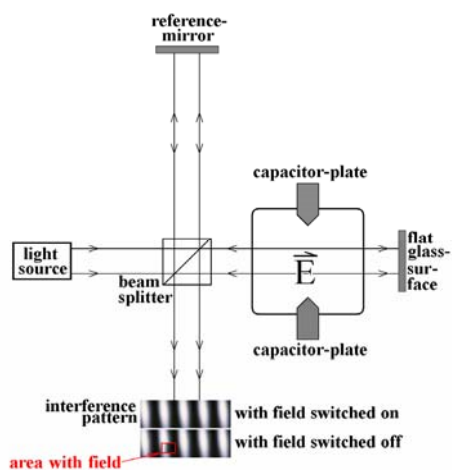

**Fig.1:**
Sketch of the principle of the white light interferometer, which was an „atos Micromap 570".
The reference-beam goes its way without being exposed to any electric field, the measuring-beam on its way to the flat smooth glass-plate can be exposed to an electric field $\vec{E}$ produced by a capacitor.
When both beams coincide, they produce an interference pattern, which will be shifted as soon as the field $\vec{E}$ is switched on – if the speed of propagation of light can be influenced by the electric field.

The interpretation of the interference pattern in fig.1 is performed according to the following explanation (see also fig. 2):
With the field switched "OFF", the interference pattern displays only stripes as they are produced by the sloped assembly of the smooth glass plate. When the field is switched "ON", it shall only be applied within the "area with field", which is marked with a red rectangle in fig.1. Consequently, the shift of the gray scale values is restricted to this specified area. Subtraction of the gray scale values of both patterns from each other deliver the shift of the values, as illustrated in fig.2.

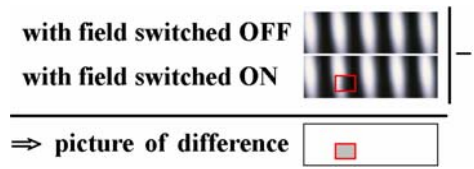

**Fig.2:** Illustration of the principle of the analysis of the interference pattern. The influence of an electric field to the speed of propagation of light can noticed as the difference of the optical path length causing the gray scale values in the difference pattern.

The measuring device of the white light interferometer is shown in Fig.3. A fastener socket with capacitor, coils and glass-plate was inserted into the white light interferometer like a specimen. Because of the limitation of the space between the objective lens and the glass-plate, this "special specimen" had to be mounted within a height of 6.5 mm (between the ground-plate of white plastic and the objective lens). A drawing of the construction is shown in Fig.3.

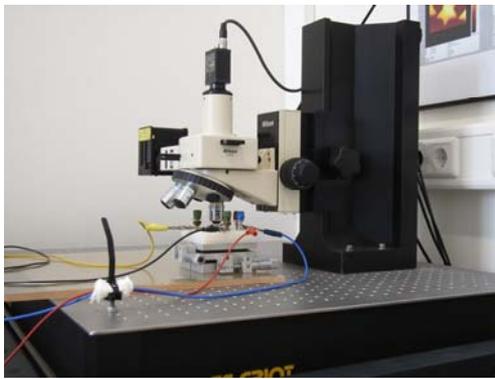

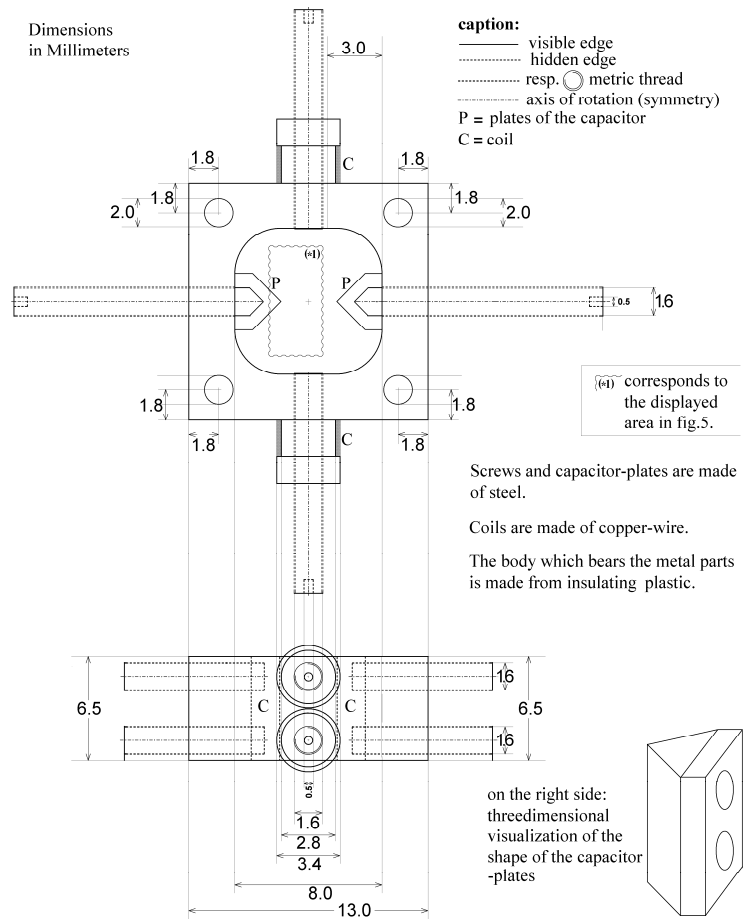

**Fig.3 (left above):**
White light interferometer with "special specimen", The ground-plate of the specimen can be identified as a white plate with electrical connections.

**Fig.4 (right side):**
Design drawing of the socket with capacitor, coils and glass-plate. It was fixed with screws on the plastic ground-plate to avoid any movement.

The capacitor for the generation of the electric field was connected to high voltage adjustable in the range of 0…13 kV. The geometry of the electric field is displayed in Fig.5.[2]

In section 3 it will be seen, that the measured data do not allow to determine exact numerical values as a result. Furthermore we shall have stray field from the top and the bottom end of the capacitor plates in mind. Nevertheless a relation between the electric field and gray scale shift seems recognizable.

[2] Special thanks go to Prof. W. Eberhard at the University of Applied Sciences Braunschweig-Wolfenbüttel for the supply of electronic equipment and to Dr.-Ing. F. Lienesch at the Physikalisch- Technische Bundesanstalt Braunschweig for the computation of the geometry of the electric field on the basis of the Finite element method.

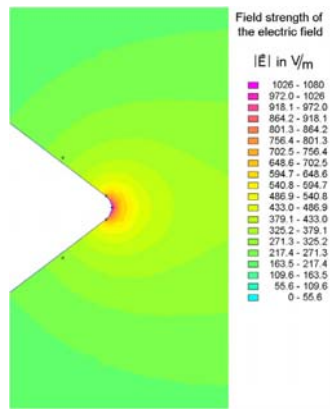

**Fig.5:**
Spatial distribution of the absolute values of the field strength generated by the capacitor plates. The picture displays a two dimensional sectional view, which is zoomed to an area similar like (∗1) in fig.4. The electric voltage between the two capacitor plates is 1.0 kV in the exemplary computation.

The coils shown in fig.4 could only produce a magnetic field up to 80 A/m at the position of the measuring-beam. This was not enough to produce a detectable influence on the propagation of the light beam. After this fact was observed, the coils were not feed any further with current, so that the magnetic field was not payed any further attention.

### 3. Measurement, analysis and discussion of the results

The central aspect of the measurements was the generation of interference patterns without and with several values of voltage between the plates of the capacitor. An example is shown in fig.6. For part (a.) capacitor plates have been with conductive connected (voltage of zero between them), for par (b.) there was a voltage of 1.56 kV between the plates. The value of the voltage has been varied from measuring-series to series as well as the slope of the glass plate, which is responsible for the density of the stripes. With the naked eye a difference between both interference patterns cannot be recognized, only a drift which is time dependant for all series of measurement, and which shall be further discussed later in the context of uncertainties of the measurements.

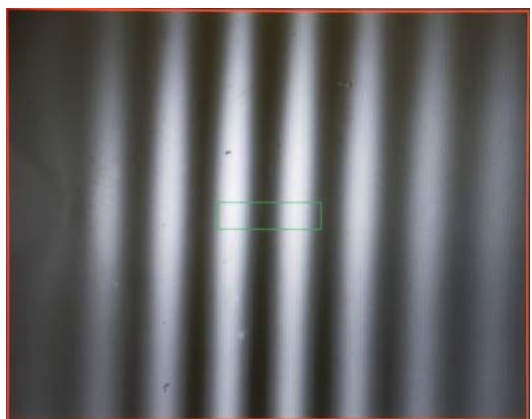
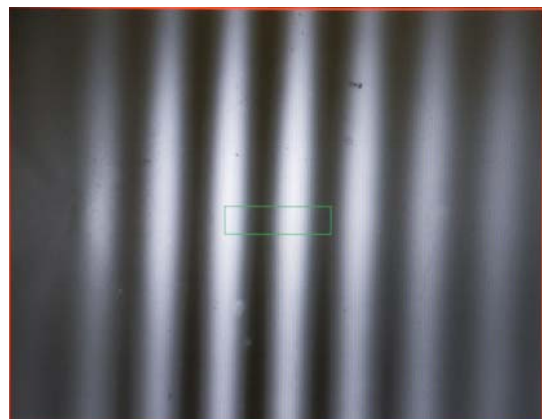

(a.) without electric field

(b.) with a voltage of 1.56 kV between the both capacitor plates

**Abb.6:**
Interference pattern with and without electric field. The left capacitor plate is located in the middle of the height at the left edge of the picture.

For the evaluation of the data the differences of the gray scale values are calculated for each pixel position. In some cases a lateral adjustment of both pictures had to be carried out, in order to compensate the lateral drift. Such an operation can be recognized as a lateral movement of the green rectangles in the middle of the pictures relatively to each other.

Fig.7 and fig.8 show typical results of difference patterns, whereas the contrast of brightness was enhanced, in order to make the perceptibility easier. Thereby the color values of „red" „green" and „blue" have been treated separately. This leads to a shift of the resultant colors in the plots.

In fig.7 the pixel-color is shifted in the very close neighbourhood to the capacitor plate, this is the region of the maximum field strength. In the middle of the picture and further on to the right side, the color values are a consequence of the slope of the glass plate and can not be interpreted in connection with the electric field of the capacitor. The voltage between the capacitor plates was 1.86 kV in the example, which leads to a maximum field strength of about 2060 V/cm.

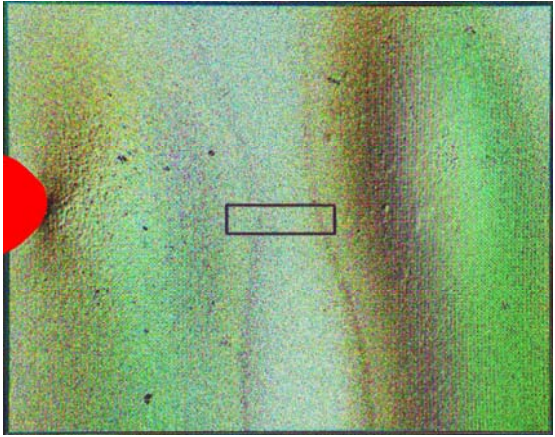

**Fig.7:**
Difference of two interference patterns with and without electric field.
The left capacitor plate is marked with red color. In its very close neighbourhood, the gray scale values are influenced. (But the shape of the influence is different from fig.5.) The influence of the gray scale values in the right side of the picture is due to the lateral shift of the pattern caused by drift, and due to some alterations in brightness when the patterns have been recorded.

When fig.7 was recorded, the flat glass plate was adjusted nearly but not perfectly perpendicular to the light beam, so that only very few interference stripes occurred. For fig.8 the glass plate was adjusted with an angle a bite more different from 90°, so that the number of interference stripes increased. In the example of the displayed picture, the capacitor plate extends into an extremum of interference. Very close to the capacitor plate, where the field strength achieve their maximum the gray scale values differ from the periodical continuation of the interference pattern as it would be without the electric field. The voltage between the capacitor plates was 1.70 kV in this example, which leads to a maximum field strength of about 1885 V/cm.

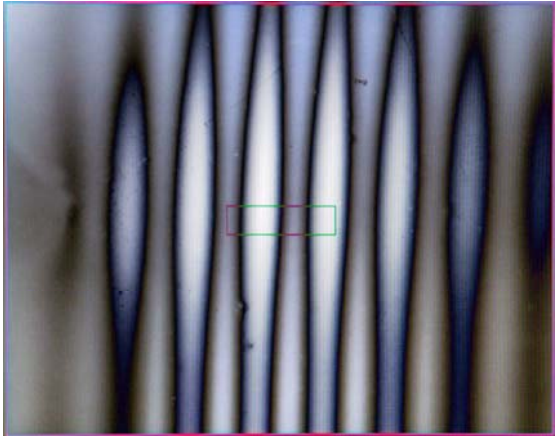

**Fig. 8:**
Difference of two interference patterns with and without electric field.
The capacitor plate is not marked with color and the contrast of the gray scale values was not changed as much as in fig.7, but the influence of the capacitor plate on the pattern can be recognized.

As stated in section 1, an optical path retardation of 520 nm corresponds to one full gray scale, which represents a color value from 0 to 256. The observed contrast covers a range of color values of about 1…20, which corresponds only to a very small fraction of a wavelength, this would be typically about $\frac{1…20}{256} \cdot \lambda$. This means that the optical path retardation can not be more than about $\Delta x = \frac{1…20}{256} \cdot 520 \ nm \approx 2...40 \ nm$. The value shall be understood only as a rough estimation of the order of magnitude of the influence of electrical fields on the speed of propagation of light, which is described in the following lines. As explained in [TUR 07] it can be said in good approximation:

$\Delta n = n - 1 = \frac{\Delta c}{c} = \frac{\Delta x}{x}$    with    $n$ = refractive index with electric field

$\Delta n$ = alteration of the refractive index caused by the electric field
$c$ = speed of light in the vacuum
$\Delta n$ = alteration of the speed of light caused by the electric field
$x$ = length of the path of light within the space of the electric field
$\Delta x$ = alteration of the light path caused by the electric field

Because the field is inhomogeneous in all spatial directions, $x$ can only be estimated to be circa $x \approx 12...15\,mm$, these are about $6.5\,mm$ for each way back and forth plus the length of stray fields. This leads to $\Delta n = \frac{\Delta x}{x} \approx \frac{2...40\,nm}{10...15\,mm} \approx 1...40 \cdot 10^{-7}$. Because of the uncertainty it is enough to say $\Delta n \approx$ few $10^{-7}$ for an electric field between 1.5 kV/cm and 2 kV/cm.

A further increase of the field strength would not be reasonable, because the experiment was performed in air, and the formation of ions should be avoided.

The uncertainties of the investigation shall be discussed in the following lines. At least we should have in mind, that it looks surprising, that an effect in the above mentioned order of magnitude, was not noticed in optics when working Kerr-cells or Pockels-cells (see [HEC 05]). Thus it is necessary to discuss possible sources of artefacts in the measurement:

● The lateral drift of the interference pattern with time (mentioned at the beginning of section 3) probably has its reason in the length-alteration of the interferometer due to temperature variations. The room had air-condition, and the temperature of thermometer was kept within $(22.2 \pm 0.2)°C$, but the solar radiation was not constant during the measuring time, and the doors of the laboratory had to be opened at least from time to time. Furthermore the assembly had to be touched with fingers and tools from time to time.

● An influence might descend from the air because of the possibility of ionization of the molecules. The ions have different refractive index than the neutral $N_2$ - and $O_2$ - molecules. To be really sure, this risk can only be excluded, if the experiment would be performed in vacuum. This will become necessary in order to obtain quantitative data in future experiments. Nevertheless in the experiment described here, the ionization of the air-molecules was avoided by a restriction of the electric field strength. The electrical breakdown of the air requires a field strength of about $10\,kV/cm...30\,kV/cm$ (depending on the humidity of the air). In the experiment the field strength was restricted to values less than $2\,kV/cm$.

By the way it should be mentioned, that fastener socket, the ground-plate and the objective lens (see fig. 2 and 3) surround the volume of the measurement which they protect to some extent against the ambient room, so that the exchange of air with the rest of the laboratory is almost suppressed. This is necessary because any movement of the air would alter the refractive index of the air by some orders of magnitude more than the electric field.

● It is imaginable that electrostatic forces might move the measuring equipment or parts of it as a consequence of electrostatic forces. In order to avoid this, the fastener socket and the ground-plate are built rather massive, and they are fixed rigidly with screws in a way that they (and even the glass-plate) could not be dislocated with a small noticeable force by hand (using a screw driver). In comparison with this, the possible electrostatic forces are much too small to cause any noticeable movement or deformation of the equipment: With regard to the dimensions of the capacitor-plates, the electrostatic charge is in the range of about $10^{-10}$C, which leads to Coulomb-forces negligible in comparison with the macroscopic forces mentioned above.

● Furthermore, a serious risk for the generation of artefacts is the fact, that the gray scale values of the photographic pictures vary from picture to picture. One of the consequences is, that the extrema of interference do not compensate each other completely when the two pictures are subtracted from each other (see for instance difference pictures in fig.7 and fig.8), as it would be expected in the ideal case of fig.2. Because if this, alterations of the gray scale values of several pixels or groups of pixels occurring in one single picture will reoccur in the difference picture. The danger is, that such gray scale artefacts might be interpreted as a part of studied effect if they occur several times or systematically. Finally this problem is the reason, why the observations reported above should not be already understood as a reliable proof for the influence of an electric field on the speed of propagation of light.

● The reliable result is: If an electric field has an influence on the speed of propagation of light, this is not more than $\frac{\Delta c}{c} \leq$ few $10^{-7}$, but it can be even much smaller than this order of magnitude.

## 4. Considerations about further investigation

Development of a theoretical model:
The measurements are based on the planning of the experiment described in [TUR 07]. The idea which led to the project originated from reflections about analogy-models such as for instance presented by [DMI 00], [DMI 02] and by [ARM 04]. However, those models are not elaborated enough that they already could produce quantitative predictions for the experiment reported here. It is desirable to get quantitative calculations from theory, based on a fundamental understanding of the process. Probably the demand will be the understanding going back to quantum theory referring to the nontrivial structure of the vacuum (see for instance review by [GIA 01]).

Optimization of the measurements:
The hypothesis of the influence of electric fields on the propagation of light sounds surprising, as well as the experimental hints for it. Thence, measurements with better precision are necessary. On the one hand they will have to help to exclude artefacts which are unnoticed up to now, on the other hand, they shall deliver reliable quantitative results. Therefore alternative measuring methods shall be discussed in the following lines.

By principle all methods to be discussed have to measure $\Delta n = \frac{\Delta x}{x}$. Consequently their resolution get better with decreasing length resolution $\Delta x$ and with increasing optical path length $x$. From this point of view it is clear, which parameters have to be optimized.

● White light interferometers have light with a coherence length as normal light bulbs or as normal daylight. In order to get coherence anyhow, the interferometer has to be built in a way that the path length of the reference-beam is identically the same as the path length of the measuring-beam within a fraction of the coherence length. Of technical reasons this is rather difficult for a long light path, such as several or many meters. The measurement reported above has a resolution of approximately $\Delta n = \frac{\Delta x}{x} \approx \frac{0.2\,nm...0.5\,nm}{12\,mm} \approx 2...5 \cdot 10^{-8}$ (according to practical experience in the laboratory). For white light interferometers it is not possible to decrease $\Delta x$ or to increase $x$ considerably. Thus, advances in resolution will probably require different methods of measurement.

● Michelson-interferometers can come to a typical $\Delta x \approx 10^{-3}...10^{-2}\lambda \approx 0.6...6\,nm$ (for HeNe-laser), depending on time and effort to optimize them. But the length of the measuring-path can be many 10 meters (or even more). An example for the resolution might be $\Delta n = \frac{\Delta x}{x} \approx \frac{0.6\,nm}{60\,m} \approx 10^{-11}$.

But this is still far from the limit (depending on time and effort), as for instance the GEO600-Experiments demonstrates [LIG 03].

● Polarimeters can be used to detect the optical path length by the means of the alteration of polarization. This leads to an accuracy in $\Delta x$ even better that in the case of a white light interferometer, whereas the measuring-path can be many 10 meters (or more).
The angular resolution of a commercially optimized polarimeter can reach the range of $\Delta \varphi \approx ...10^{-3°}...$ . If we put this into relation with the angle of a full circle ($\varphi = 360°$), it is possible to achieve a length resolution $\Delta x$ of: $\frac{\Delta x}{\lambda} = \frac{\Delta \varphi}{\varphi} = \frac{...10^{-3°}...}{360°} \Rightarrow \Delta x \approx \left(...3 \cdot 10^{-6}...\right) \cdot \lambda \approx ...2 \cdot 10^{-12}...\,m$.

In combination with a long measuring-path, an example for the resolution of $\Delta n$ could be like $\Delta n = \frac{\Delta x}{x} \approx \frac{...2 \cdot 10^{-12}...\,m}{60\,m} \approx ...4 \cdot 10^{-14}...$, this is roughly the range of $\Delta n = \frac{\Delta x}{x} \approx ...10^{-14}...$.
(The optimization also depends strongly on time and effort.)
The idea is to use a splitted beam (see fig.9). A reference-beam will not be exposed to an electric field, for a measuring-beam the field can be toggle-switched "on" and "off". After bringing both beams together, the polarization will depend on the optical path length of the measuring-beam.

Annotation: The influence of the optical elements on the polarisation was not taken into account in fig.9.

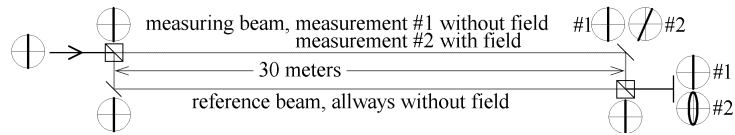

**Fig.9:** Sketch of an imaginalbe pola-rimeter-experiment.

## Literature

[ARM 04]    Ether theory of gravitation: why and how ?
            Mayeul Arminjon, 2004, arXiv:gr-qc/0401021 v1

[BIA 70]    Nonlinear Effects in Quantum Electrodynamics. Photon Propagation and Photon Splitting in an external Field.
            Z. Bialynicka-Birula und I. Bialynicki-Birula, 1970  Phys. Rev. D, Vol.2, No.10, p.2341

[DMI 00]    Mechanical Analogies for the Lorentz Gauge, particles and antiparticles
            Valery P. Dmitriyev, 2000, arXiv:physics/9904049

[DMI 02]    Gravitation and electromagnetism
            Valery P. Dmitriyev, 2002, arXiv:physics/0207091

[FOM 76]    Measurements of the solar gravitational deflection of radio waves in agreement with general relativity.
            E. B. Fomalont und R. A. Sramek, 1976, Phys. Rev. Lett. 36, p. 1475

[FOM 77]    The deflection of radio waves by the sun.
            E. B. Fomalont, 1977, Comments in Astrophysics 7, p. 19

[GIA 01]    Field correlators in QCD. Theory and applications.
            A. Di Giacomo, H. G. Dosch, V. I. Shevchenko and Yu. A. Simonov, 2001
            arXiv:hep-ph/0007223 v2

[GOE 96]    Einführung in die spezielle allgemeinen Relativitätstheorie
            H. Goenner, 1996, Spektrum Akademischer Verlag, ISBN 3-86025-333-6

[HEC 05]    Optik,
            Eugene Hecht, 2005, Oldenbourg-Verlag, ISBN3-486-27359-0

[LIG 03]    Detector Description and Performance for the First Coincidence Observations between LIGO and GEO
            The LIGO Scientific Collaboration, 2003, arXiv:gr-qc/0308043 v3

[KLÜ 60]    The determination of Einstein's light deflection in the gravitational field of the Sun." von
            H. v. Klüber, 1960 in "Vistas in Astronomy", p. 47, Pergamon Press, Editor A. Beer

[RIK 00]    Magnetoelectric birefringences of the quantum vacuum
            G. L. J. A. Rikken and C. Rizzo, 2000, Phys. Rev. A, Vol.63, 012107

[RIZ 07]    The BMV project: Biréfringence Magnétique du Vide"
            Presentation by Carlo Rizzo, März 2007, see   http://moriond.in2p3.fr/J07/sched07.html

[SCH 02]    Gravitation
            U. E. Schröder, 2002, Verlag Harri Deutsch, ISBN 3-8171-1679-9

[TUR 07]    A Hypothesis regarding the Refraction of Light and the precise planning of an Experiment for its Verification.
            Claus W. Turtur, März 2007, arXiv:physics/0703271v1

## Author's Adress:

Prof. Dr. Claus W. Turtur
University of Applied Sciences Braunschweig-Wolfenbüttel
Salzdahlumer Straße 46 / 48
Germany - 38302 Wolfenbüttel
Email: c-w.turtur@fh-wolfenbuettel.de
Tel.: (++49) 5331 / 939 – 3412